\documentclass[showpacs,amsmath,
twocolumn,
aps,prl]{revtex4}
\usepackage{graphicx}
\usepackage{dcolumn}

\newcommand{\ket}[1]{\ensuremath{\left|#1\right\rangle}}
\newcommand{\bra}[1]{\ensuremath{\left\langle#1\right|}}
\newcommand{\unit}[1]{\ensuremath{\, \mathrm{#1}}}

\begin{document}

\title{A clock containing a massive object in a superposition of states; 
what makes Penrosian wavefunction collapse tick?}

\author{Tjerk H. Oosterkamp and Jan Zaanen}
\affiliation{Leiden Institute of Physics, Niels Bohrweg 2, 2333 CA  Leiden, The Netherlands}




\begin{abstract}
Penrose has been advocating the view that the collapse of the wave function is rooted in the incompatibility between general relativity and quantum mechanics. On the basis of conceptual analysis, he arrived at an estimate for the collapse time. 
To better understand his estimate, in this paper we present a thought experiment, which singles out the role of time-dilations in massive superpositions. First we investigate the behavior of a hypothetical clock containing a component which can be in a superposition of states. 
The clock contains a massive object, whose only purpose is to introduce a curvature of space time into the problem. We find that a state of this massive object with a smaller radius, but with the same mass, experiences a larger time dilation. 
Considering a coherent superposition of the large and small object, introduces an ambiguity in the definition of a common time for both states. We assert that this time ambiguity can be thought to affect the time evolution of a state in different ways and that the relative phase difference between these different interpretations can be calculated. We postulate that the wave function collapse will occur when this phase difference becomes of order unity. An absolute energy scale enters this equation and we recover Penrose's estimate for the 
collapse time by equating the absolute energy scale to the rest mass of the object. 
\end{abstract}
\maketitle

\section{1. Introduction}
The plan of this paper is as follows.  Penrose \cite{Penrose} has put forward an approach to estimate the time scale at which gravity will start to play a role in the quantum mechanical time evolution of heavy objects, which we discuss in detail in section 2. Throughout the paper we call these states `Schr\"odinger's cat states', which in this paper we will assume to be spherical. In our approach, we argue that the gravitational effect, which will eventually destroy their coherent superposition, revolves around gravitational time dilations which are unequal for 
the Schr\"odinger's cat states. 
In sections 3 and 4 we discuss three different ingredients that we combine in section 5, where we arrive at an estimate of the timescale at which gravity will start to play a role in quantum mechanics.
Because of the similarity of the final result, we postulate that our approach may in essence be the same as Penrose's and that in Penrose's conceptual analysis the precise role of these time dilations may be included implicitly. 
While Penrose's analysis is fully in the Newtonian limit, we introduce the concept of the superposition of time dilations, which goes beyond the Newtonian limit. However, the superpositions we consider do not have relative motion and in this sense remain Newtonian \cite{Penrose_Newtonian_limit}.

Utilizing Einstein's heuristic principle of a clock, in section 3 we present a novel gedanken experiment tailored to make it easy to track the progress of time explicitly, inspired by a famous GR experiment based on an  analysis by Shapiro. We find that the the dilation differences due to micrometer sized spheres are tiny. 

In section 4 we discuss the ambiguity in time which arises if a wavefunction describes a sphere in a superposition of two different diameters. This consideration demonstrates that this ambiguity interferes with the unitary time evolution of a coherent superposition. We then arrive  at a criterium showing when this Einsteinian ambiguity of time becomes noticeable.  This turns out to require 
an absolute energy and, taking the rest mass of the object for this energy, we recover Penrose's dimensional estimate for the collapse time.  
 
In section 5 we refine the 
argument by considering continuous mass distributions, also highlighting the role of Penrose's assertion that one state in the coherent superposition encounters the space time curvature of its quantum partner, which in our approach takes the form of one state encountering the clock of its quantum partner.  
 
\section{2. The Penrose view on wavefunction collapse.}
The precise status of the collapse of the wavefunction in a quantum measurement has been a highly contentious and
confusing subject since its introduction some 80 years ago. In our personal experience, the majority view is to assert that there is in fact no problem at all, invoking the "for all practical purposes" idea that 
conventional decoherence due to the interaction with the environment suffices.
However,  there is still quite some dissent in the form of a variety of ideas  claiming that the 
collapse cannot be explained on this basis. This  invokes alternative interpretations varying from quite mystical (e.g., that human consciousness is the culprit) up to the quite practical
"objective collapse" ideas. The latter assert that there is just new physics at work, of a kind that is in principle measurable, while it can eventually be comprehended as 
a reasonable physical process which does not invoke human observers, many worlds, or whatever.  

This school of thought has, at the least, the benefit of a prediction -- it is not mere philosophy. It is an empirical fact that microscopic objects like electrons or quarks fully submit
to the unitary world of orthodox quantum physics, while macroscopic things like cars are never found in coherent superposition. Given the hypothesis that the collapse
is a measurable physical process, it should take place in a regime of scale between the microscopic and the macroscopic. Especially the former has been pushed upwards, by
the demonstration that quasi-macroscopic objects like flux qubits \cite{flux qubit} do not collapse within the bath-decoherence time scales that can, at present, be achieved in the laboratory. 
Such an "objective collapse", however, does need to happen when things get quite big. 

What determines this scale? While others have suggested to use gravity inspired models \cite{Karolyházy, Diósi, Ghirardi}, we depart from the fact that there is only one proposal available on the basis of known physics: the idea of Penrose that the wave function is based on the incompatibility of
the unitary time evolution which is at the heart of quantum physics, and the space time of general relativity \cite{Penrose}. This conflict is manifest in all attempts to get a grip on quantum gravity. 
The bottom line is that unitarity, the time evolution governed by linear transformations in Hilbert space with the Hamiltonian as generator of time translations, is not a diffeomorphic invariant. 

As stressed by Penrose, unitarity requires a global time like Killing vector \cite{Penrose}, and this becomes an issue when gravitationally inequivalent space times are involved in a coherent 
quantum superposition. In principle, a point like identification between space times with different mass distributions is an impossibility according to general relativity. 
This in turn is an issue when one is dealing with simple Schr\"odinger's cat states, since the live and the dead cat will have a different mass distribution making it impossible 
to assign a global time like direction in a space time that can be "shared" by the two cats in the superposition. Henceforth, Schr\"odinger's cats face in principle the same 
problem as black holes. In the regime of particle physics down to molecular physics, well removed from both the Planck scale and the macroscopic scale, 
gravity is so weak that this cannot possibly play any 
role. Therefore, unitarity is just fine in the realms of atoms up to the tera electron volts of the best particle accelerators. But on the human scale, gravity becomes noticeable:
could it be that the collapse occurs because gravity wins, destroying unitary evolution and thereby causing the collapse? 
This is the key question posed by Penrose \cite{Penrose}. 

At present, any insight in the microscopic theory that would lead to the "gravity wins" outcome  is lacking, and in the absence of theoretical guidance, all that remains is dimensional analysis. 
Penrose suggested that a "Planck scale" can be identified, associated with the gravitational wave functional collapse that lies in the regime 
in between microscopics and macroscopics, based on natural dimensions of quantum physics and gravity.  Planck's constant $\hbar$ is obviously the quantity associated 
with quantum physics, carrying the dimension of energy times time.  This is a convenient dimension to convert energy into time, and Penrose \cite{Penrose}  asserts that the time
associated with the wave function collapse is given by,

\begin{equation}
\tau_G = \frac{\hbar}{\Sigma_G}
\label{tauG}
\end{equation}

where $\Sigma_G$ is a gravitational quantity with the dimension of energy associated with the inequivalence of space times encountered in Schr\"odinger's
cat like situations.  He then suggested that this should be a sort of relative self-energy, based on the difference in two mass distributions.  The cat is surely 
non-relativistic and also is in a regime where gravity is weak, and therefore one should look in the Newtonian limit. The "alive cat" defines a gravitational potential well,
associated with its mass distribution. The gravitational self-energy is defined by keeping this potential fixed, while one computes the gravitational energy associated
with moving the mass distribution to become coincident with its "dead cat" quantum copy,

\begin{eqnarray}
\Sigma_G & = &  \frac{1}{G} \int d^3\mathbf{x} (\bf{f_a}-\bf{f_b})\cdot(\bf{f_a}-\bf{f_b}) \nonumber  \\
& = &  \int d^3\mathbf{x} [\Phi_a-\Phi_b][\rho_a-\rho_b]
\label{gravselfen}
\end{eqnarray}

where $\bf{f_a}$ and $\bf{f_b}$ are the vectors indicating the strength of the gravitational fields associated with two different mass distributions, $\rho_a$ and $\rho_b$, that are in superposition with each other. $\Phi_a$ and $\Phi_b$ are the gravitational potentials associated with these mass distributions, $d^3\mathbf{x}$ indicates an integral over the three spatial dimensions and $G$ is Newton's constant.

Assuming that the cats correspond with simple spherical masses, $M$, with radius $a$ displaced over a length $d$ where $a << d$, one arrives 
at an estimate for the order of magnitude of the gravitational collapse time \cite{Penrose} 

\begin{equation}
\tau_G = \frac{5}{6}  \frac{ \hbar a} {G M^2} 
\label{penrosedim}
\end{equation}

Intriguingly, one finds that for a "cat" of typical size $a=1\unit{ \mu m}$, which has a weight in the range of micrometer sized bacteria ($10^{-15} \unit{kg}$) and which is in superposition with itself after being displaced by a length of
$b=1\unit{ \mu m}$, $\frac{M^2G}{a}=6.6*10^{-35} \unit{J}$ and it takes a time $\tau_G$ of a few seconds to collapse its wavefunction.  This is precisely in the range, which has not been explored experimentally. It is however
quite appealing for experimentalists, since there is a serious potential that this regime comes into reach using the latest technology, e.g. \cite{ClelandMartinis, Bouwmeester, vanWezelOosterkamp}.   At the same time, this estimate has been  criticised merely on basis that any effect of gravity on time and so forth should be so minute that it can 
be completely ignored \cite{GtHooft}.  After all, the dogma is that gravity and quantum physics 
should only clash at the conventional Planck scale. Our main result is that we will arrive at a rational explanation why this intuition might be in principle misleading.  

\section{3. The clocks of Schr\"odinger's cat:  A Shapiro type gedanken experiment}

Even for the purpose of dimensional analysis, Penrose's estimate for the gravitational wave function collapse time is ad-hoc. The assertion that gravity enters via the gravitational self energy,
is not rooted in a detailed consideration of how the "ambiguity of time in the superposition" arises in general relativity. 
Instead, Penrose argues that the gravitational energy is the only quantity 
he can identify in this Newtonian regime which relates to the superposition of mass distributions, while it can be balanced with $\hbar$ to yield a reasonable scale. 

We wish to point out here that it is in fact quite straightforward to address this ambiguity of space time as it arises in gravity. We employ Einstein's favorite heuristic method 
of tracking how clocks tick in the reference frames of observers traveling with the Schr\"odinger's cat quantum copies.  Since their mass distributions are different,
the clocks attached to the quantum copies will indicate a different time in a classic GR manner, and it is obvious that this disagreement should correspond with the 
time ambiguity as of relevance to the destruction of the unitary time evolution.

In the following we calculate for micrometer sized spheres what the scale of this effect is. At stake is that, following Penrose, the two quantum copies are characterized by a different sense of time relative to each other. To discern to what extent the sense of time is different in these two space-times, we need a measure that can be shared by both 'universes'. 
We propose a rather natural clock that can accomplish the goal of measuring the relative difference in the sense of time in this situation. 
Later in this paper we estimate the ambiguity in the phase evolution of a mass distribution using two different clocks: one for each reference frame of the two different Schr\''odinger's cat quantum copies. 
Note that unlike previous papers, which investigate time dilation effects on quantum mechanics due to a single gravitational potential \cite{Pikovski, Helfer, Anastopoulos, Zych, Lammerzahl, Dimopoulos} we look at the difference between two time dilation effects of two superposed states. We wish to make it clear however, that we do not derive an expression which provides the new time evolution which ensues in the presence of a superposition of time dilation effects.

In our approach, the gravitational side gives rise to an ambiguity in a quantity with the dimension of time (instead of energy as in Penrose's estimate) and we then need to work out how this enters the quantum mechanical equation. We will argue that this
involves necessarily an {\em absolute} energy scale. In section \ref{On the absolute energy}, we take for this energy the relativistic rest mass of the "cat".
In section \ref{Continuous mass distributions} we no longer take a single time for each quantum copy of the cat, which depends on the mass distribution, but we consider the situation in which every atom making up the cat can be assigned its own 'clock'. 
With this we can quantify the phase ambiguity, which the ambiguity of time creates between two possible time evolutions. We find that the result of our approach then becomes coincident with Penrose's estimate.

To start our considerations, we will consider a Gedanken experiment. It is given in by convenience, since it allows us to use material from GR textbooks to estimate 
the ticking of the clocks. Although we are not aware of any physical principle prohibiting the construction of the device, the barriers to overcome in order to make it work 
in the laboratory might well be insurmountable. However, we just employ it in order to make the calculations easier and we expect it to be trustworthy in the limited
sense of getting the order of magnitude at which time becomes ambiguous right.  

It consists of a  ball made of a material having the property that it undergoes a zero temperature (quantum) 
first order phase transition where the volume of the material drastically changes. This is less exotic than it might appear. The lanthanides metals cerium, praseodymium and 
gadolinium as well as the actinide plutonium show a thermal "volume
collapse" transition \cite{rare-earths} where the volume of these metals can decrease by as much as 15\%, as related to a drastic, cooperative change of their f-electron systems 
from a delocalized- to a localized nature. For recent work in Ce see Ref. \cite{Cerium}. A first challenge for the 
material scientist is to drive such a transition to very low temperatures.  In principle it is possible to force the ball in a coherent superposition of its large- and small 
volume phases right at the zero temperature transition. Subsequently, the ball has to be 
kept isolated from the environment in order to prohibit decoherence, while it surely has to be kept at a very low temperature. To accomplish this experimentally could well be an impossible pursuit, but in principle it might be done.

\begin{figure}[htb]
\centering
\includegraphics[width=5in]{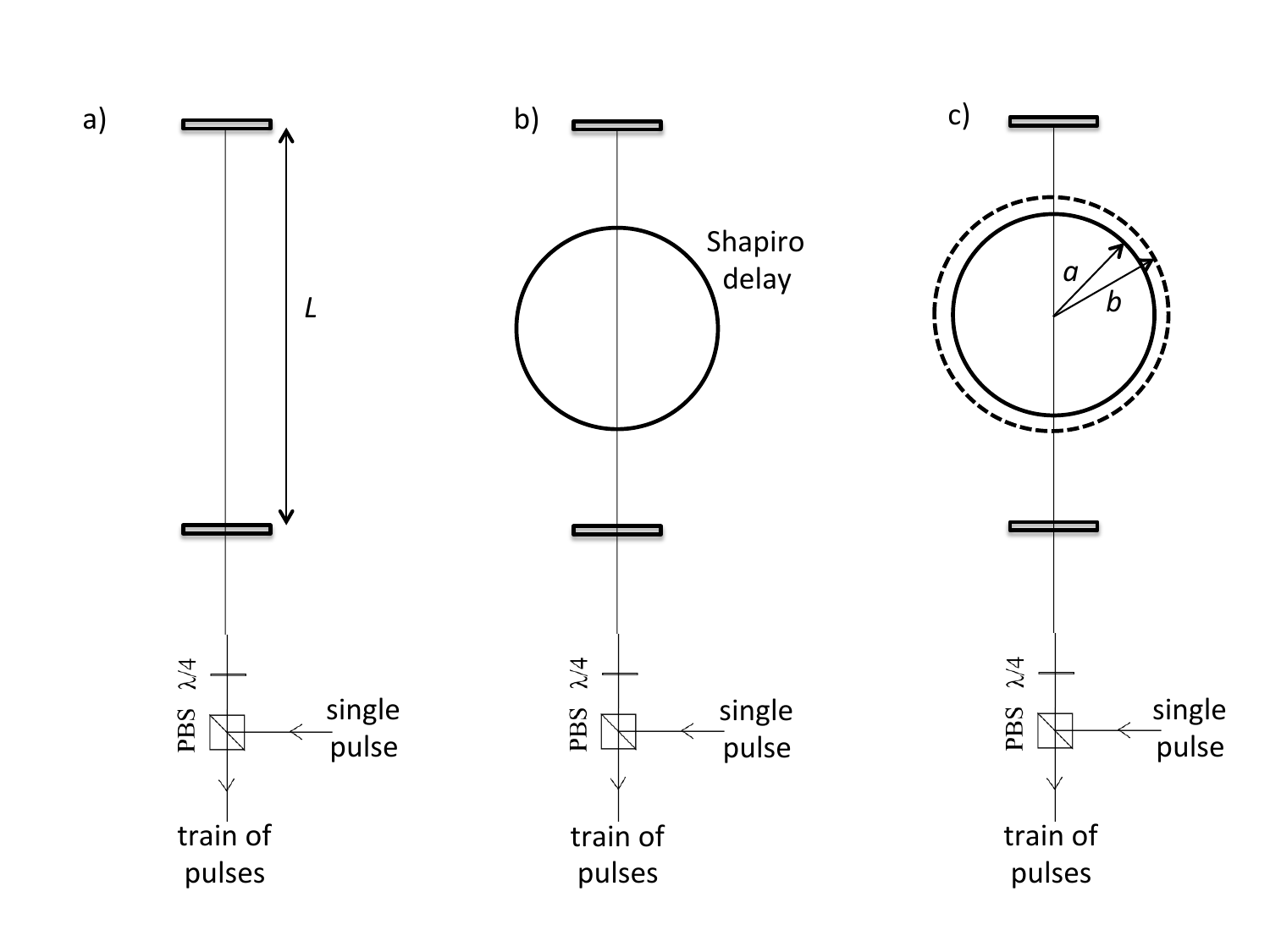}
\caption{three clocks, each consisting of a strong laser pulse coupled into a cavity through a polarizing beam splitter and a $\lambda/4$-plate, and subsequently bouncing around in a cavity. The lower mirror has a slight transmission, such that a train of pulses is coming out of the cavity. Panel a) and b) show a clock without and with a mass inside the cavity, while panel c) shows a clock in which the mass is in a superposition of states.}
\label{three clocks}
\end{figure}


Why do we introduce this ball in a superposition of its two "volume states"?  The reason is that we can directly apply a famous GR story, as related to the direct 
measurement of gravitational time dilation: the Shapiro time delay effects, as tested in the 1960's exploiting the solar system \cite{Shapiro_PRL}. By reflecting off a radar pulse 
of the surface of the planet Venus, its time of flight was measured. It turned out that as Venus passes behind the sun, the radio pulse experiences a delay of approximately $200 \unit{\mu s}$, due to the space time curvature induced by the mass of the sun.  One already 
anticipates that such time dilation effects will be quite delicate when dealing with objects of the weight of E. coli.    

For the sole purpose of measuring the times associated with the mass distributions of the large and small ball, we present in Fig. 1 three different clocks. Panel a) shows a conventional clock, while the other two clocks contain a massive object, which for panel c) is in a superposition of volume states. The conventional clock in panel a) consists of a laser pulse bouncing between two mirrors separated by a distance $L$. The end mirror at the top of the image is perfectly reflective, while the entrance mirror is almost perfectly reflective but has a small transmission $\epsilon$.  The light enters the cavity from a laser $I$ which emits a single gaussian shaped pulse $p(t)=\frac{1}{\epsilon^2}exp(-t^2/t^2_{pulse})$ centered around $t=0$, p(t) of monochromatic light with wavelength $\lambda$ and pulse length $ t_{pulse}<<L/c$, where $c$ is the velocity of light.  The light exits the cavity as a train of equally spaced pulses, $p_{clock}(t)= \sum_{n>0}  p(t-n\Delta t)$, separated by $\Delta t = 2L/c$. Note that these light pulses do not necessarily need to be detected.

\begin{figure}[htbp]
\centering
\includegraphics[width=\columnwidth]{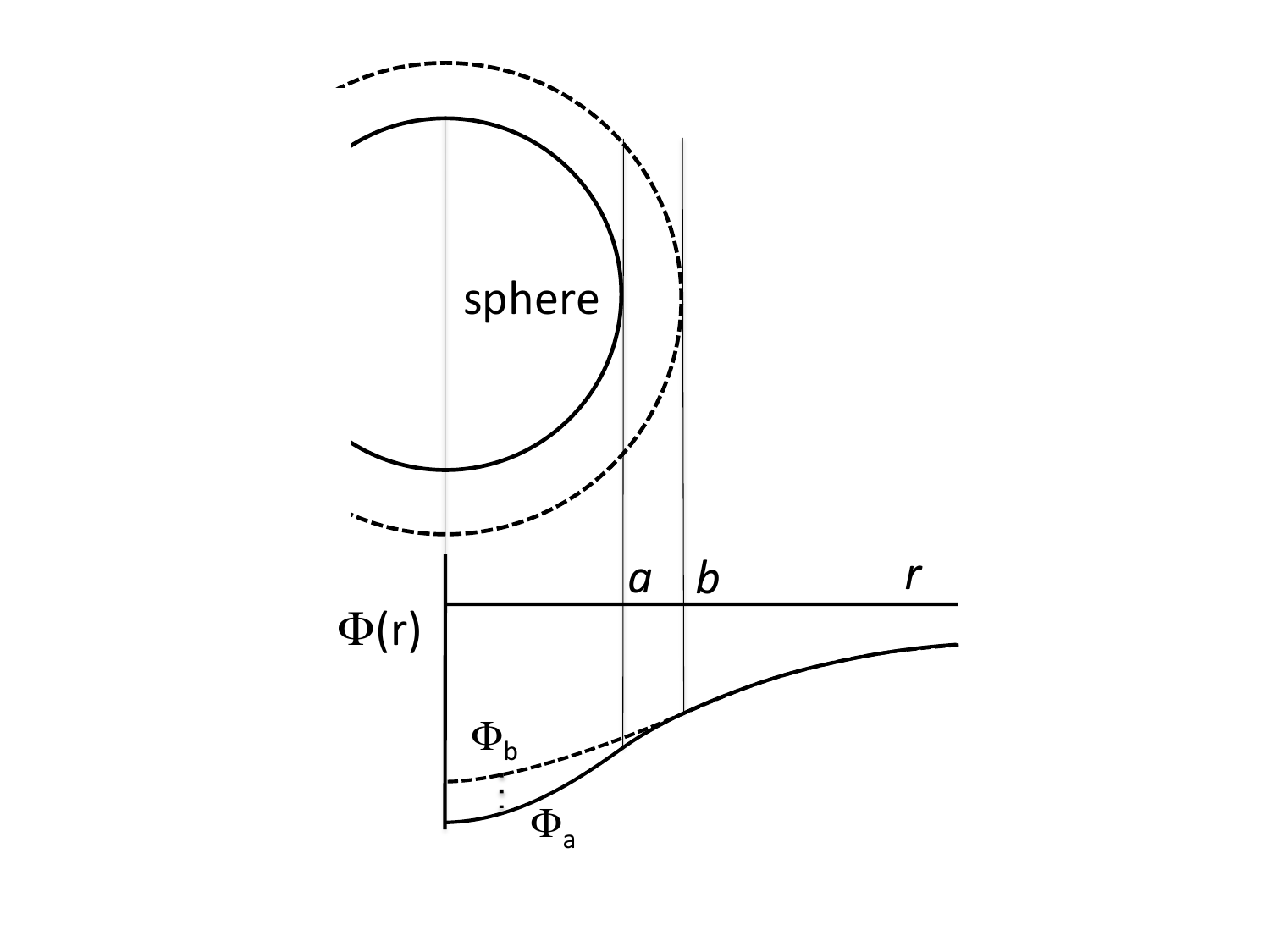}
\caption{The gravitational potential near a sphere as a function of the distance from the center of the sphere $r$. The solid and dashed curves indicate the potential of a sphere of radius $a$ and of radius $b$, respectively. The dotted vertical line illustrates that, for lack of a theory of quantum gravitation, we don't know the gravitational potential of a sphere in a superposition. The vertical dotted line indicates that the potential at a certain radial position of two states with radius $a$ and radius $b$ might be anywhere between the two solid curves and, similarly, the time dilation might be governed by a gravitational potential anywhere in between these two values.}
\label{three potentials}
\end{figure}

Panel b) shows a clock, which is influenced by the gravitational field due to the presence of a mass. A spherical mass, $M$, resides within the cavity, whose sole purpose is to introduce a space-time curvature, and since the measurement of time is just a Gedanken experiment, the sphere can be regarded as perfectly transparent while it does not affect the laser beam in any way, other than changing the space time curvature. 
Due to the space-time curvature induced by the sphere, the Shapiro delay \cite{Shapiro_PRL} is expected to occur, with our spherical mass taking the role of the sun. 

In figure \ref{three potentials}, we plot the gravitational potential $\Phi(r)$, from which the gravitational force, $F_G$, felt by a test mass, $M_{test}$, can be derived through $F_G=-M_{test}\frac{d}{dr} \Phi(r)$. Outside a sphere of radius $a$ the potential is given by $\Phi(r)=-GM/r$ with $G$ Newton's constant and $r$ the distance from the center of the sphere, while inside the sphere it shows a quadratic behaviour $\Phi(r)=-(GM/a) (3/2-r^2/2a^2)$, since the gravitational force grows linearly inside the sphere. 
For a light beam running through the center of the sphere, it is straightforward to integrate the gravitational time dilation.
The magnitude of this effective time delay, for two passes, from the bottom mirror to the end mirror at the top of the diagram and back, $\Delta T_a=T_0-T_a$, follows from a consideration involving the effect of the gravitational potential on the light traveling in the radial direction. Here $T_0=\frac{2L}{c}$ is the time for two passes when there is no mass present and $T_a$, is the time for two passes when a mass $M$ with radius $a$ is present in between the mirrors.


\begin{equation}
 \Delta T_a = - \frac{4}{c^3}\int^{L/2}_{-L/2} \Phi(r) dr
\end{equation}

This would imply that the ticking of time, $t_a$, measured using the period $T_a$ of our clock with a sphere with radius $a$ (Fig. 1b), would be slightly slower than the ticking of time, $t$, measured using the period $T_0$, of the same clock without the sphere (Fig. 1a):

\begin{equation}
T_a=T_0- \Delta T_a = (1-\epsilon_a) T_0
\end{equation}

\begin{equation}
t_a = (1-\epsilon_a) t
 \label{timedelay}
\end{equation}

with 

\begin{equation}
\epsilon_a= \frac{\Delta T_a}{T_0}= -\frac{2}{Lc^2} \int_{-L/2}^{L/2} \Phi_a(r) dr
\label{epsilon}
\end{equation}
and $T_0=2L/c$. 

Let us now compute the time difference associated with a laser pulse traveling through the two different volume states of the sphere, characterized by the radii $a$ and $b$.

\begin{equation}
 \Delta T_a - \Delta T_b=- \frac{8}{c^3} \int^{L/2}_0 (\Phi_a(r)-\Phi_b(r) ) dr 
\end{equation}

\begin{equation}
 = -\frac{8GM}{c^3} (\int^{a}_0 [\frac{3}{2a}-\frac{r^2}{2a^3}]dr -\int^{b}_0 [\frac{3}{2b}-\frac{r^2}{2b^3}]dr - \int^{b}_a \frac{dr}{r} )
\end{equation}
and since the first two integrals cancel only the last one remains:
\begin{equation}
 \Delta T_a -\Delta T_b= \frac{8GM}{c^3} \ln( \frac{b}{a}) \approx \frac{8GM}{c^3}\frac{b-a}{a}
 \label{Deltatijden}
\end{equation}

where the last step is possible if $a$ is only slightly smaller than $b$. 

A sphere with a radius of $b=5 \unit{\mu m}$, engineered to have a low temperature volume phase transition of 15\% with a density of $5000 \unit{kg/m^3}$, will have a typical mass $M=3*10^{-12} \unit{kg}$. If a superposition with its low volume state were to be achieved, $a/b=0.95$, and we were to take a distance between the mirrors surrounding the spheres of $L=10 \unit{\mu m}$, the time difference of a laser pulse passing these two mass distributions would come out as $\Delta T_a - \Delta T_b=  2.6* 10^{-48} \unit{s}$. and the dimensionless parameter

\begin{equation}
\epsilon_a-\epsilon_b=\frac{\Delta T_a-\Delta T_b}{T_0}= 4*10^{-35}
\end{equation}

This is indeed an exceedingly small effect. It just confirms the intuition that one would not expect relativistic time dilation effects to play any role on the scale of biological 
cells, given that the sun was barely heavy enough to measure the Shapiro delay using 1960's radar technology. 

\section{4. On the time ambiguity and the absolute energy of Schr\"odinger's cat.}
\label{On the absolute energy}
Importantly, we have now calculated differences in time, as would be inferred from light-signals, for an observer external to the spheres. In section \ref{Continuous mass distributions} we estimate the effect of the difference in time experienced by the two spheres in their different volume states (rather than the time experienced by an external observer), by giving each atom in the sphere their own clock. After all, clocks that would be constructed around different small parts of the sphere would tick differently, because the gravitational potential is not constant within a sphere itself. 
However, before we do that, for simplicity we discuss how this 
ambiguity in time arises and plays out in the time evolution of the coherent superposition of the two spheres in the case where each sphere is assumed to have a single time which depends only on the radius of the sphere. Given that this time ambiguity is a small number, and we have only the modest task of identifying after how much time this small number might have an effect, 
we can proceed in a perturbative fashion. Let us first ignore these time delay effects to specify the Schr\"odinger's cat state involving the large- and small volume states
of the sphere $\ket{a}$ and $\ket{b}$. We prepare the state at time zero in the superposition  $\ket{\psi(t=0)}=\alpha \ket{a}+\beta \ket{b}$ and allowing for slightly different
energies $E_a$ and $E_b$ we find that this state will have evolved after a time $t$ in, 

\begin{equation}
\ket{\psi(t)} =\alpha e^{-\frac{i}{\hbar} E_a t }\ket{a} + \beta  e^{-\frac{i}{\hbar} E_b t }\ket{b} 
\end{equation}

We learned in the previous section that because of the gravitational time dilation effect the two states actually experience a different
time, since their gravitational potentials are different!

We assert that this time ambiguity introduced by general relativity will be the key ingredient that can introduce non-unitary effects into quantum mechanics. If we would want to find something that can explain why we see wavefunction collapse when performing a measurement, we need a non-unitary effect and it is this effect that may provide the clue we need. 
The point of the matter is that while a pure state $\ket{\psi} =\ket{a}$ may have a time evolution according to time $t_a$, and likewise a pure state $\ket{\psi} =\ket{b}$ may have a time evolution according to time $t_b$, we have no answer as to what time we should consider for the time-evolution of $\ket{a}$ if it is part of $  \ket{\psi(t)} =\alpha \ket{a} + \beta  \ket{b}  $. We would like to emphasize here that, if it is indeed the case that the time-ambiguity causes the time-evolution of a state $\ket{a}$  to change in the presence of a component of the wavefunction $\beta \ket{b}$ with $\beta \neq 0$ and where $\ket{b}$ has a different mass distribution than $\ket{a}$, we may indeed have identified a source of non-unitarity.
  
Proceeding naively, pretending that the superposition
is still subjected to a unitary evolution, we might instead expect the state,

\begin{equation}
\ket{\psi(t)} =\alpha \hspace{1mm}e^{-\frac{i}{\hbar} E_a t _a}\ket{a} + \beta \hspace{1mm}  e^{-\frac{i}{\hbar} E_b t_b }\ket{b} 
\end{equation}
where $t_a$ and $t_b$ are the 'Shapiro' times introduced in Eqn. \ref{timedelay} and computed in the previous section.
Alternatively, if one insists that the time evolution should be governed by a single time, one might attempt to write down a time evolution such as:
\begin{equation}
\ket{\psi(t)} =\alpha \hspace{1mm} e^{-\frac{i}{\hbar} E_a (\lvert \alpha \rvert^2t _a+\lvert \beta \rvert^2t _b)}\ket{a} + \beta \hspace{1mm} e^{-\frac{i}{\hbar} E_b (\lvert \alpha \rvert^2t _a+\lvert \beta \rvert^2t _b) }\ket{b} 
\label{weighed_times_in_evolution}
\end{equation}
where the average of the two times is taken. This would leave us with the problem, however, that the time evolution of the sum of two states is no longer the sum of the time evolutions of the separate states as is the case in ordinary quantum mechanics.

Neither of these time evolutions is according to the rules that quantum mechanics poses. However, these equations do express in a minimal way that the quantum time evolution is affected by the ambiguity of relativistic time. Although we do not know how to solve the ambiguity of time, now we do have a way to express that due to the superposition of states, there is in principle room for non-unitarity to enter the time evolution. 

In this paper we are merely looking for a time-scale at which  the end of unitary quantum mechanics might become noticeable in an experiment. Surely the last thing we expect to appear are the familiar Rabi oscillations, since these are rooted in the assertion that the time evolution is unitary.
Neither do we expect to arrive at ordinary decoherence. Decoherence is the process in which the density matrix $\rho(t) = \ket{\psi(t)} \bra{\psi(t)}$ becomes  a diagonal matrix when tracing away microscopic degrees of freedom and this is a result of perfectly unitary behavior. Wavefunction collapse, on the other hand, results in one of the diagonal elements becoming unity, and turning all the other diagonal elements of $\rho(t)$ equal to zero, which requires a manifestly non-unitary process.

We use the ambiguity of time introduced above for the purpose of identifying the time-scale after which the ambiguity of time might have a manifestation in an experiment. Of course one is not supposed to use equations of this kind once the ambiguity of time has become manifest.  But even though we don't know how to write down the actual time evolution, we can calculate when the difference between the possible ways the phases might reach an amount of order unity. 
We therefore proceed by investigating the phase difference, rather than speculating about the proper way to write down the time evolution.
The simplest phase ambiguity that we might look at is:
\begin{equation}
 \phi (t) =  -\frac{1}{\hbar} ( E_a t_a-  E_b t_b) 
 \label{simple}
\end{equation}
with the two times $t_a$ and $t_b$ parametrized in terms of the Shapiro delay parameters $\epsilon_a$ and $\epsilon_b$ of the two mass distributions as  
$t_a=(1-\epsilon_a) t$ and $t_b=(1-\epsilon_b) t$ associated with the states $\ket{a}$ and $\ket{b}$, respectively. 

The phase ambiguity which at the end of this paper will turn out to reproduce the Penrose timescale, is the following: 
 \begin{equation}
 \phi (t) = -  \frac{1}{\hbar} E_a (t_a - t_b)  + E_b (t_b-t_a) = - \frac{1}{\hbar} (E_a - E_b) (t_a-t_b) 
 \label{penrose_phase_difference}
\end{equation}
This phase difference expresses how much the phases in front of $\ket{a}$ and $\ket{b}$ vary, if one were to assume that the time used for the time evolution changes, depending of the weights $\alpha$ and $\beta$. I.e., if the state $\ket{\psi}$ were completely $\ket{a}$ one would take $t_a$ and if the $\ket{\psi}$ were completely $\ket{b}$ one would take $t_b$. But in between, as $\lvert \beta \rvert^2$ is increased from $0$ to $1$ the time is ambiguous. One might say that the ambiguity that we are trying to quantify is the effect introduced by the other space time, as expressed in a minimal way in Eqn. \ref{weighed_times_in_evolution}, which for state $\ket{a}$ can be as large as $t_a-t_b$  and for state $\ket{b}$ can be as large as $t_b-t_a$.

For simplicity, and because we first would like to point out another essential ingredient (the absolute energy scale of a Schr\''odinger's cat), we first investigate Eqn. \ref{simple}, which can be rewritten as 
\begin{equation}
\phi(t) =-\frac{1}{\hbar} ((E_a -E_b)t - (\epsilon_a E_a- \epsilon_b E_b)t)
\label{penrose_phase_difference2}
\end{equation}
although later, at the end of section \ref{Continuous mass distributions}, we will use the phase difference of \ref{penrose_phase_difference}.

Looking at either of these\ two phase differences reveals an interesting surprise: in the presence of the gravitational ambiguity of time, the time evolution
of the wave function {\em becomes sensitive to the absolute value of the energy of the states involved.}  
In normal quantum mechanics, i.e. when the two states in superposition 
agree on their time, only the energy differences $E_a - E_b$ matter, because the mean energy $E_{mean} = ( E_a +E_b )/2$ appears only in the  overall phase 
$i E_{mean} t$, which is therefore pure gauge and devoid of physical implications.

But when the time depends on the state, as in our Shapiro situation, it does matter what the mean energy is. Below we evaluate Eqn.\ref{penrose_phase_difference2} for two energies $E_{a}^{\star}$ and $E_{b}^{\star}$ that are close to each other but at a large absolute energy  $\Delta E \gg E_a , E_b$:

\begin{equation}
E_{a}^{\star}=\Delta E + E_a    
\qquad\text{and}\qquad
  E_{b}^{\star}=\Delta E + E_b      
\end{equation}

After this substitution, the phase shift $\phi(t)$ becomes $\phi^{\star}(t)$,

\begin{equation}
 \phi^{\star} (t) =  \phi(t)  - \frac{t}{\hbar}  (\epsilon_a-\epsilon_b)\Delta E  
\end{equation}

We are now facing a question which is unusual in quantum physics, for which a quite natural answer is found in general relativity: what 
to take for the absolute energy $\Delta E$? Of course the relativist's answer would be $\Delta E= M c^2$, the rest mass of the sphere. Inserting this and using the expression found for $\epsilon_a-\epsilon_b$ for our particular geometry, we obtain,  

\begin{equation}
 \phi^{\star} (t) =  \phi(t)  + \frac{t}{\hbar} \frac{8GM}{c^2}\frac{b-a}{2La} * Mc^2
\end{equation}

\begin{equation}
 \phi^{\star} (t) =\frac{t}{\hbar}  8GM^2(\frac{b-a}{2La}) 
\end{equation}

where in the last step $\phi(t)$ has been neglected because $Mc^2>>E_a, E_b$. Note that the term $c^2$ in the numerator of the expressions for $\epsilon_a-\epsilon_b$ has been canceled by the multiplication with the energy $Mc^2$ and that we do not take into account gravitational energy contributions because the gravitational field is extremely weak.

If we now insert the values taken for $L=10 \unit{\mu m}$, $a=4.75 \unit{\mu m}$, $b=5 \unit{\mu m}$ and $M=2.6*10^{-12} \unit{kg}$ in the previous section, we find that after a time of $t=75 \unit{\mu s}$ the phase $\phi^{\star}$ has reached a value of $2\pi$. This timescale is very much like the scale that Penrose arrived at with his analysis.

Although our Gedanken experiment departs from a Newtonian, non-relativistic limit we would like to point out that the above expression is invariant under a Lorentz boost. The phase difference discussed here, should not depend on whether or not the experiment is observed by a spectator which moves in a frame with a velocity $v$ relative to our 'cat'. This is an issue, especially since we have to explicitly invoke the relativistic energy $\Delta E$ to arrive at the above expression.
The expression $(\epsilon_a-\epsilon_b)\Delta E = \frac{\Delta T_a-\Delta T_b}{T_0} \Delta E$ is Lorentz invariant because to an outside observer, who sees the clock fly by at a velocity $v$, the energy $\Delta E=Mc^2$ increases to $\Delta E^{\prime}=\Delta E / \sqrt{1-\frac{v^2}{c^2}}$, while the clock without the masses will appear to tick more slowly, $T_0$ becoming $T_0^{\prime}=T_0  / \sqrt{1-\frac{v^2}{c^2}}$, thereby decreasing the value of $\epsilon_a-\epsilon_b$. 
These two effects exactly cancel because $T_0$ is in the denominator. Finally, note that the expression for $\Delta T_a-\Delta T_b$ contains the gravitational potential $\Phi(r)$, which is not a Lorentz scalar. In the limit where the logarithm $\ln{\frac{b}{a}}$ in Eqn. \ref{Deltatijden} can be written as $\frac{b-a}{a}$ both lengths $b-a$ and $a$  in the fraction will be Lorentz contracted in the same way and thus the expression for $\phi^{\star}$ is Lorentz invariant. 

\section{5. Continuous mass distributions: atoms acquiring individual clocks.}
\label{Continuous mass distributions}
Let us now generalize our heuristic clock model: we will argue that our line of thought may lead to the same collapse time as the one proposed by Penrose \cite{Penrose}. While Penrose balances the gravitational self energy with $\hbar$ to arrive at an estimate of a characteristic time after which quantum superposition will collapse $\tau_G = \frac{\hbar}{\Sigma_G}$, we point out that $\phi^{\star}$ grows with time and that the collapse of the wavefunction might occur by the time $\phi^{\star}=2\pi$, which for our example of the superposition of two states consisting of concentric spheres of different densities which for simplicity we both gave a single time, such as described in the previous section, would result in a collapse time

\begin{equation}
\tau_G  = \frac{4 \pi  \hbar L a }{8 G M^2 (b-a)}
\end{equation}

which indeed contains all the dimensions in the expression one would arrive at when evaluating the integral for the gravitational self energy, called on by Penrose. Of course, the length of our clock does not appear in the integral for the gravitational self energy, nor do we think that the length of our clock should be an ingredient that is fundamental in our analysis.

The way in which we can do away with the length of the clock,  allowing us to write our approach in the same integral form as Penrose, is by giving each atom making up the sphere its own clock. This is natural, realizing that it is rather unsatisfying that our first model depends on the details of how the clock is constructed, such as the separation of the mirrors $L$ or whether one chooses to send the light beam right through the center of the sphere or through a path that is off center. We therefore proceed by giving each atom in the sphere its own clock, to ask subsequently which phase shift the different sphere states pick up when we parametrize the time $t$ experienced by each separate atom.

We consider a solid body where the system of atoms has a spontaneously broken translation symmetry. This implies that the state of the sphere with radius $a$ ($\ket{a}$) is a product state of all the nuclei in the sphere:
\begin{equation}
\ket{a}= \prod_j \ket{r_{j,a}}
\end{equation}
where $\ket{r_j}$ are real space wave-packets for  the $j$th nucleus which is part of the sphere localized in its crystal position, and the subscript $a$ serves as a reminder that all positions $r_{j,a}$ make up the lattice of the sphere with radius $a$.

We would like to write down the gradual build up of a phase difference between two states with different mass distributions. We start out with the time evolution of a single state.

\begin{equation}
\ket{a(t)}= \prod_j   e^{-\frac{i}{\hbar} m_j c^2  t_{j,a} } \ket{r_{j,a}} = e^{-\frac{i}{\hbar} \sum_j m_j c^2  t_{j,a}}  \prod_j   \ket{r_{j,a}} 
\end{equation}
where $m_j$ denotes the mass of the $j$th atom. This is perfectly sound when all $t_{j,a}=t$. We have now deviated from the quantum mechanical rules, however, by asserting that each atom carries its own local time, denoted by the subscripts in $t_{j,a}$. With the gravitational potential $\Phi(r_{j,a})$, which each atom feels at its position this leads to  

\begin{equation}
t_{j,a}=(1-\epsilon_{j,a})t
\end{equation} 
with 

\begin{equation}
\epsilon_{j,a}=-\frac{2\Phi_a(r_{j,a})}{c^2}
\label{infinitesimal_eps}
\end{equation}

which now is now determined by the configuration of all atoms making up the sphere with radius $a$.

Using the same parametrization of the Shapiro like time delay, we arrive at a phase ambiguity,  $\phi_{diff,a-b}$, between the pure states $\ket{\psi}=\ket{a}$ and $\ket{\psi}=\ket{b}$, analogous to the phase ambiguity of Equation  \ref{simple}

\begin{equation}
\phi_{diff,a-b}=-\frac{1}{\hbar}\left(\sum_j m_j c^2  t_{j,a}-\sum_j m_j c^2  t_{j,b}\right)
\end{equation}

which becomes, after substituting $t_{j,a}=\left(1+\frac{2\Phi_a(r_{j,a})}{c^2}\right)t$ and $t_{j,b}=\left(1+\frac{2\Phi_b(r_{j,b})}{c^2}\right)t$, where the extra subscript added to the spatial coordinate $r_{j}$ serves as a reminder that each atom has a different coordinate in state $\ket{a}$ versus state $\ket{b}$:

\begin{equation}
\phi_{diff,a-b}=-\frac{t}{\hbar} [\sum_j m_j 2 \Phi_a(r_{j,a}) - \sum_j m_j 2\Phi_b(r_{j,b})]
\end{equation}

which may also be written in integral form as

\begin{equation}
\phi_{diff,a-b}=-\frac{2t}{\hbar}  \int d^3\mathbf{x} [\Phi_a(\mathbf{x}) \rho_a(\mathbf{x})-\Phi_b(\mathbf{x}) \rho_b(\mathbf{x})]
\end{equation}
where $\rho_a(\mathbf{x})$ and $\rho_b(\mathbf{x})$ are the mass distributions of the sphere with radius $a$ and radius $b$, respectively.

\begin{figure}[htbp]
\begin{center}
\includegraphics[width=\columnwidth]{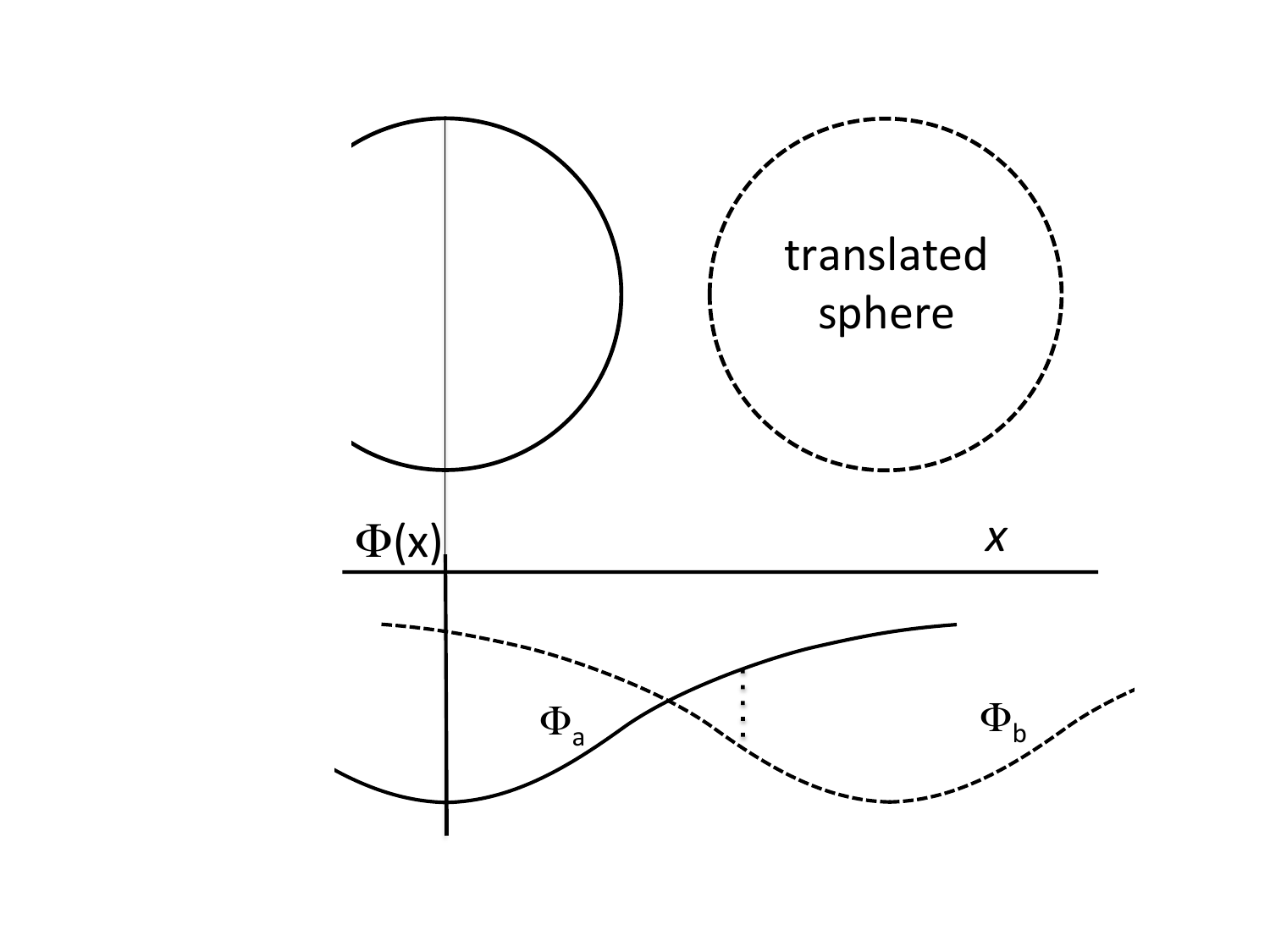}
\end{center}
\caption{The gravitational potential near a sphere as a function of position measured along an axis through the center of two spheres, $x$. The solid and dashed curves indicate the potential of a sphere of the same radius, but displaced in x. The vertical dotted line indicates that the time dilation experienced by an atom that is part of a sphere that is in a superposition of states of two different positions might be determined by a gravitational potential anywhere in between $\Phi_a$ or $\Phi_b$.}
\label{displaced potentials}
\end{figure}

Note that it should be understood that the gravitational potential is taken to be smooth at the atomic scale, because we have assumed in our derivation that the gravitational potential is taken to have a single value per atom. In fact this was used to arrive from Eqn. (\ref{epsilon}) at Eqn.(\ref{infinitesimal_eps}).

Now, when asking at which time $t_G$ the phase ambiguity analogous to Equation \ref{simple}, accrued between states $a$ and $b$ reaches $\phi_{diff,a-b}=2\pi$, one arrives at 

\begin{equation}
\tau_G=\frac{\hbar}{4\pi \int d^3\mathbf{x} [\Phi_a(\mathbf{x}) \rho_a(\mathbf{x})-\Phi_b(\mathbf{x}) \rho_b(\mathbf{x})]}
\label{coltimefinal}
\end{equation}

which closely resembles the integral form of the collapse time proposed by Penrose equations \ref{tauG} and \ref{gravselfen}, containing two of the four terms in the integral $ \int d^3\mathbf{x} (\Phi-\Phi^{\prime})(\rho-\rho^{\prime}) $ of Eq. \ref{gravselfen}.

This is due to the fact that, in the above, we have only calculated the phase difference between the two {\em separate} product  states and we have not tried to answer what the actual time evolution of a superposition of states will be, nor what the ambiguity of the gravitational potential would be when it is produced by a superposition of states. For instance, considering
the particular case of  a sphere in superposition of two states, which are only {\em displaced} over some distance rather than stretched uniformly in all directions, such as to get a different density (see Fig. \ref{displaced potentials}), 
the calculated phase difference from equation Eq. (\ref{coltimefinal}) would vanish altogether. This is quite unsatisfying since displaced spheres are surely the most elementary way to ask the gravitational wave function collapse 
question in a Schr\"odinger's cat type setting, and we are clearly still missing an ingredient.  
As pointed out previously in Eqn. (\ref{weighed_times_in_evolution}) and in the discussion after Eqn. (\ref{penrose_phase_difference}) , instead of equating the phase difference of Eqn. (\ref{simple}) to $2 \pi$, one could also choose to equate the phase ambiguity of Eqn. (\ref{penrose_phase_difference}) to $2 \pi$.
We now state that the reason is as follows: Penrose has invoked a kind of relative self energy, based on the difference in two mass distributions, which is quadratic in the differences in mass distribution. Our re-interpretation in terms of superposition of clocks introducing an ambiguity in the time used to express the ordinary quantum time evolutions is essentially the same. In the case of a coherent superposition of states
like $\ket{\psi}=\alpha\ket{a}+\beta\ket{b}$, it is implicit in the gravitational self energy construction that the maximum time dilation ambiguity experienced by state $\ket{a}$ is actually associated with the space time determined by the mass distribution of its quantum copy $\ket{b}$ and the other way around. 
Therefore in the case of the "Shapiro times" the collapse time is associated with the  
 {\em difference} between the two gravitational fields $\Phi_a-\Phi_b$, indicated by the vertical dotted lines in Figure \ref{three potentials} and \ref{displaced potentials}. This yields,

\begin{equation}
\tau_G=\frac{\hbar}{4\pi \int d^3\mathbf{x} [\Phi_a(\mathbf{x})-\Phi_b(\mathbf{x})][ \rho_a(\mathbf{x})- \rho_b(\mathbf{x})]}
\end{equation}

This expression is now exactly the same as Penrose's expression for the characteristic time he associated with wavefunction collapse when combining Eqn. \ref{tauG} and \ref{gravselfen}.

\section{6. Conclusion.}
Surely the power of Penrose's logic is to lift the quantum measurement debate from the philosophical- to the empirical realms. The question whether 
objective state reduction is the one that is chosen by nature, and the issue whether gravity or quantum physics is loosing out in the real theory of 
quantum gravity, is only decidable by experiment.  All we have accomplished is to construct a simple rational explanation why the ad-hoc dimensional
analysis of Penrose can make sense after all. Since the Shapiro experiment was successful, it is evident that our estimate for the gravitational time ambiguity
is physical. Also the way that this enters into the time evolution of the quantum superposition is very elementary. Although natural from the point of view of
dimensions, the main outcome of our analysis is that in order for the gravitational collapse to happen on the "E. coli scale" the absolute energy scale that 
is required in quantum mechanics, when time gets ambiguous, has to be the relativistic rest mass. The consequence of a successful gravitational collapse
experiment would therefore be that when unitarity comes to an end, the rest mass of Schr\"odinger's cat is no longer a quantity that can be gauged away. 
We hope that this will be a guidance for those theorists searching for the "gravity first" theory of quantum gravity.  
\vskip6pt

\maketitle

\enlargethispage{20pt}



THO conceived the gedanken experiment, JZ conceived the lanthanides-metal zero temperature phase transition and the other elements are shared. Both authors gave final approval for publication. 


THO thanks the Foundation for Fundamental Research on Matter (FOM) and the Netherlands Organisation for Scientific Research (NWO) for financial support.

We thank Jasper van Wezel, Koenraad Schalm, Dirk Bouwmeester, Roger Penrose and Ana Ach\'ucarro for discussions.



\end{document}